\documentclass[fleqn,twoside]{article}
\usepackage{espcrc2}

\usepackage{graphicx}
\usepackage[figuresright]{rotating}

\newcommand{\AmS}{{\protect\the\textfont2
  A\kern-.1667em\lower.5ex\hbox{M}\kern-.125emS}}

\newcommand{\eq}[1]{{\frenchspacing Eq.~(\ref{#1})}}
\newcommand{\fig}[1]{{\frenchspacing Fig.~(\ref{#1})}}

\newcommand{\beq}{\begin{equation}}
\newcommand{\eeq}{\end{equation}}

\title{The $\eta '$ signal from partially quenched Wilson fermions }

\author{Harmut Neff\address[iasa]{Institute of Accelerating Systems \& Applications, P.O.\ Box 17214, 10024 Athens, Greece}%
        \thanks{Electronic mail: neff@theorie.physik.uni-wuppertal.de},
        Th.\ Lippert\address[uniw]{Dept.\ of Physics, University of Wuppertal, 42097 Wuppertal, Germany},
        J.W.\ Negele\address{Center for Theoretical Physics, MIT, Cambridge, Massachusetts 02139, USA},
        K.\ Schilling\addressmark[uniw],
        }
       
\begin{document}
\begin{abstract}
  We present new results from our ongoing study of flavor singlet
  pseudoscalar mesons in QCD. Our approach is based on (a) performing
  truncated eigenmode expansions for the hairpin diagram and (b)
  incorporating the ground state contribution for the connected meson
  propagator.  First, we explain how the computations can be
  substantially improved by even-odd preconditioning. We extend
  previous results on early mass plateauing in the $\eta'$ channel of
  two-flavor full QCD with degenerate sea and valence quarks to the
  partially quenched situation.  We find that early mass plateau
  formation persists in the partially quenched situation.

    \vspace{1pc}
\end{abstract}

\maketitle

\section{Introduction}
Hadron spectroscopy has been the testbed for monitoring the progress
of lattice QCD calculations for over more than two decades.  The
$\eta'$ meson has been rather evasive, however
\cite{Venkataraman:1997xi,AliKhan:1999zi,Michael:2001rv,Neff:2001zr,Schilling:2002gm}. 
Being a flavor singlet object, its
correlation function is a combination of a connected (octet)
contribution and the infamous hairpin diagram.  The latter is
difficult to evaluate, for it is the correlator of two trace
expressions over the inverse hermitean Dirac operator, $Q^{-1}$.

So far we have been computing $D$ by means of a truncated eigenmode
representation  of $Q^{-1}$,
\begin{equation}\label{eq:eo_full} 
Q^{-1}= \sum_{i=1}^N \frac{1}{\lambda_i} | \psi_i \rangle \langle \psi_i | \; ,
\end{equation}
with $Q | \psi_i \rangle =\lambda_i | \psi_i \rangle$ and
$N=\mbox{dim}(Q)$.  'Truncated' means that we restrict the sum
\eq{eq:eo_full} to $i \ll N$, i.e.\ to some low lying eigenmodes
\cite{Neff:2001zr}. Another important ingredient of our approach is a
ground state projection of the connected correlator (one loop ground
state analysis $=$ OLGA), prior to the combination with the data for the
hairpin-operator.

For the $16^3 \times 32$ SESAM lattice at $\beta =5.6$ we showed for 2
different quark mass values that the 300 lowest lying eigenmodes of
$Q$ suffice to approximate $D$ well enough.  With OLGA, we found long
$\eta'$ mass plateaus with onset at the very first time slice.  The
results appear to be fully consistent with former stochastic estimator
analyses, which provides evidence that truncation effects from
\eq{eq:eo_full} in the SESAM operating conditions \cite{Neff:2001zr}
can be safely discarded.

We observed in Ref. \cite{Neff:2001zr} that stochastic estimator
techniques and TEA require comparable computational effort.
Actually, we can reduce the costs of TEA by means of preconditioning.

\section{Even-odd preconditioning of $Q$}

Since the Wilson Dirac matrix $M$ connects nearest neighbor sites
only, it can be readily even-odd preconditioned, i.e.  cast into the
form
\begin{equation}
\tilde{M}= 
\left( \begin{array}{cc}
1 &  0 \\
0& 1- \kappa^2 D_{oe} D_{eo} \\
\end{array} \right) \; ,
\end{equation}
where $e (o)$ stands for even(odd).  Note that $\gamma_5 (1- \kappa^2 D_{oe}
D_{eo})$ is a hermitean matrix with orthogonal eigenvectors $\phi_i$
and real eigenvalues $\sigma_i$. In terms of these eigenmodes,
$\tilde{Q}^{-1}=(\gamma_5 \tilde{M})^{-1}$ reduces to
\begin{equation}
\tilde{Q}^{-1}= \sum_{i=1}^{N/2} \frac{1}{\sigma_i}
\left| \left. \begin{array}{c} 0 \\ \phi_i \end{array} \right>
\left<  \begin{array}{c} 0 \\ \phi_i \end{array} \right.
\right| \; .
\end{equation}
For brevity we omitted a traceless term $\gamma_5 {\boldmath
  1}_{\mbox{\rm\tiny even}}$.

\eq{eq:eo_full} now becomes
\begin{equation}\label{eq:eo_pre}
{\mbox{Tr}}\, 
Q^{-1}= 
{\mbox{Tr}} 
\sum_{i=1}^{N/2} \frac{1}{\sigma_i}
\left| \left. \begin{array}{c} \kappa D_{eo} \phi_i  \\ \phi_i \end{array} \right>
\left<  \begin{array}{c}  \kappa D_{eo} \phi_i  \\ \phi_i \end{array} \right.
\right|  \; .
\end{equation}

The quark loop signals as determined from the two different sets of
vectors, i.e.\ from \eq{eq:eo_full} and \eq{eq:eo_pre}, are
practically indistinguishable. Note that
$\tilde{M}$ offers two crucial benefits: 1. it requires only half the
memory and 2.  the eigenmode algorithm applied on $\gamma_5 (1-
\kappa^2 D_{oe} D_{eo})$ is improved by a factor 4 in convergence rate.
In fact the order of the Chebyshev (accelerating) polynomials can be
cut down from 80 to 20, see \cite{Neff:2001zr}.

\section{Unquenched 2-flavor $\eta'$ mass analysis}

Let us present next an update of our recent unquenched 2-flavor
$\eta'$ mass analysis \cite{Neff:2001zr}.  We have by now completed
the quark loop computations at 4 different sea quark mass values, each
gauge field ensemble containing ${\cal O}(200)$ gauge fields.

\begin{figure}[!htb]
\includegraphics[width=5cm,angle=270]{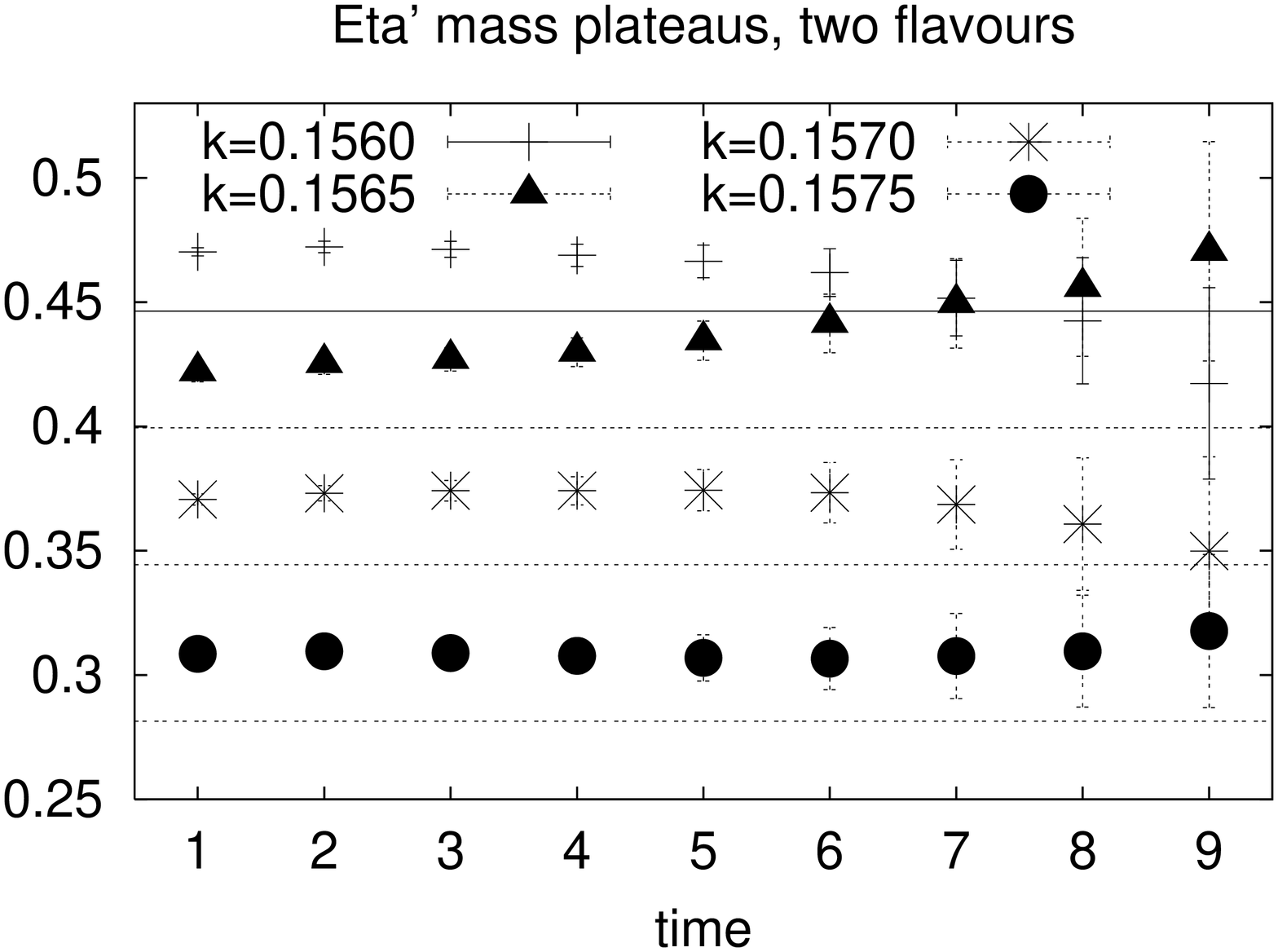}
\vskip-.7cm
\caption{$\eta'$ mass plateaus from unquenched
  QCD with two degenerate flavors, plotted in lattice units, for 4
  values of $\kappa_{sea}$. The lines correspond to the octet masses.}
\label{fig:mass_plateaus}
\vskip-.3cm
\includegraphics[width=5cm,angle=270]{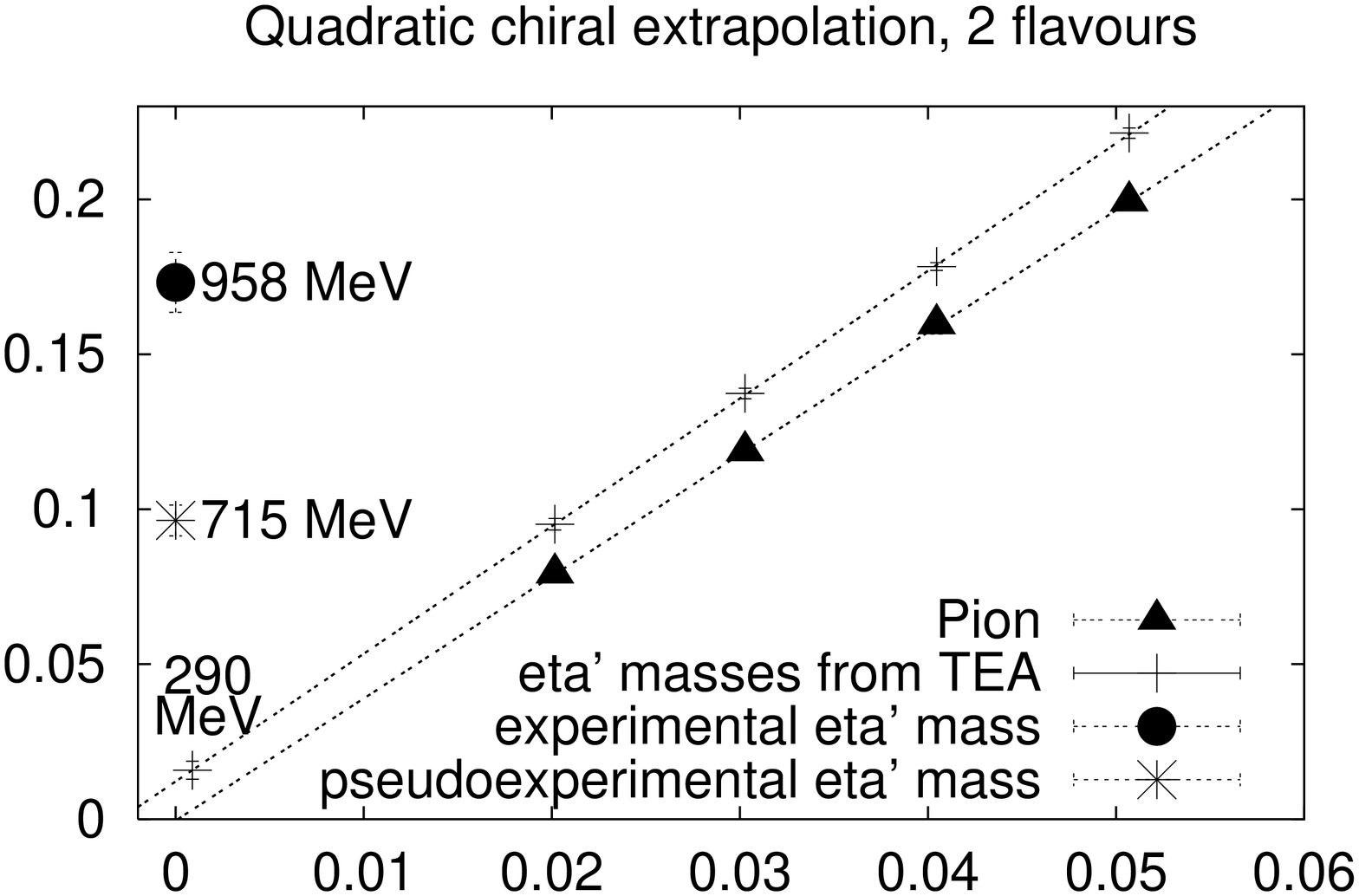}
\vskip-.3cm
\includegraphics[width=5cm,angle=270]{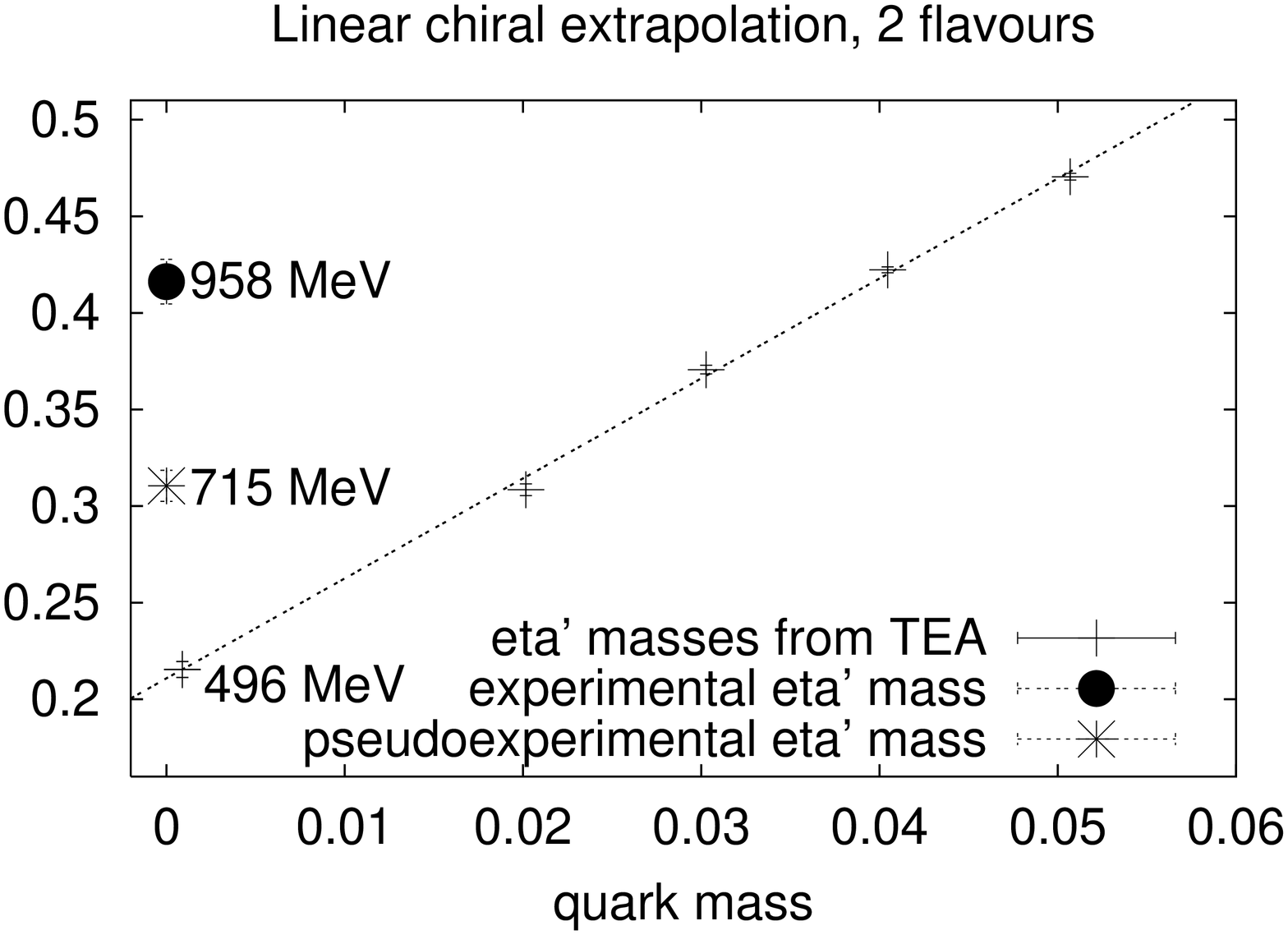}
\vskip-.5cm
\caption{
  Chiral extrapolations of the pseudoscalar octet and singlet masses
  in unquenched QCD.}
\label{fig:chiral_ext}
\vskip-.6cm
\end{figure}

The local $\eta'$ masses as extracted from logarithmic derivatives of
the $\eta'$-correlator are exhibited in \fig{fig:mass_plateaus}.  We
find that OLGA produces long and stable mass plateaus.  As expected
the quality of the plateaus improves when going more chiral in sea
quark masses.

In \fig{fig:chiral_ext} we plot the chiral extrapolations
for the singlet masses. Our data clearly favor the quadratic fit,
which yields $\chi^2=0.14$, compared to $\chi^2=5.5$ for the linear
extrapolation.

We retrieve a definitely non vanishing mass gap between singlet and
nonsinglet masses after quadratic chiral extrapolation.  The value of the
computed mass gap is far from the experimental one (even if we scale the
$\eta'$-mass according to the Witten-Veneziano expectation down to a
pseudoexperimental value as indicated in \fig{fig:chiral_ext}). 
Physically this could be due either to limitations of unimproved Wilson
fermions or the omission of strange quarks. Here we address the latter.

\begin{figure}[!htb]
\includegraphics[width=5cm,angle=270]{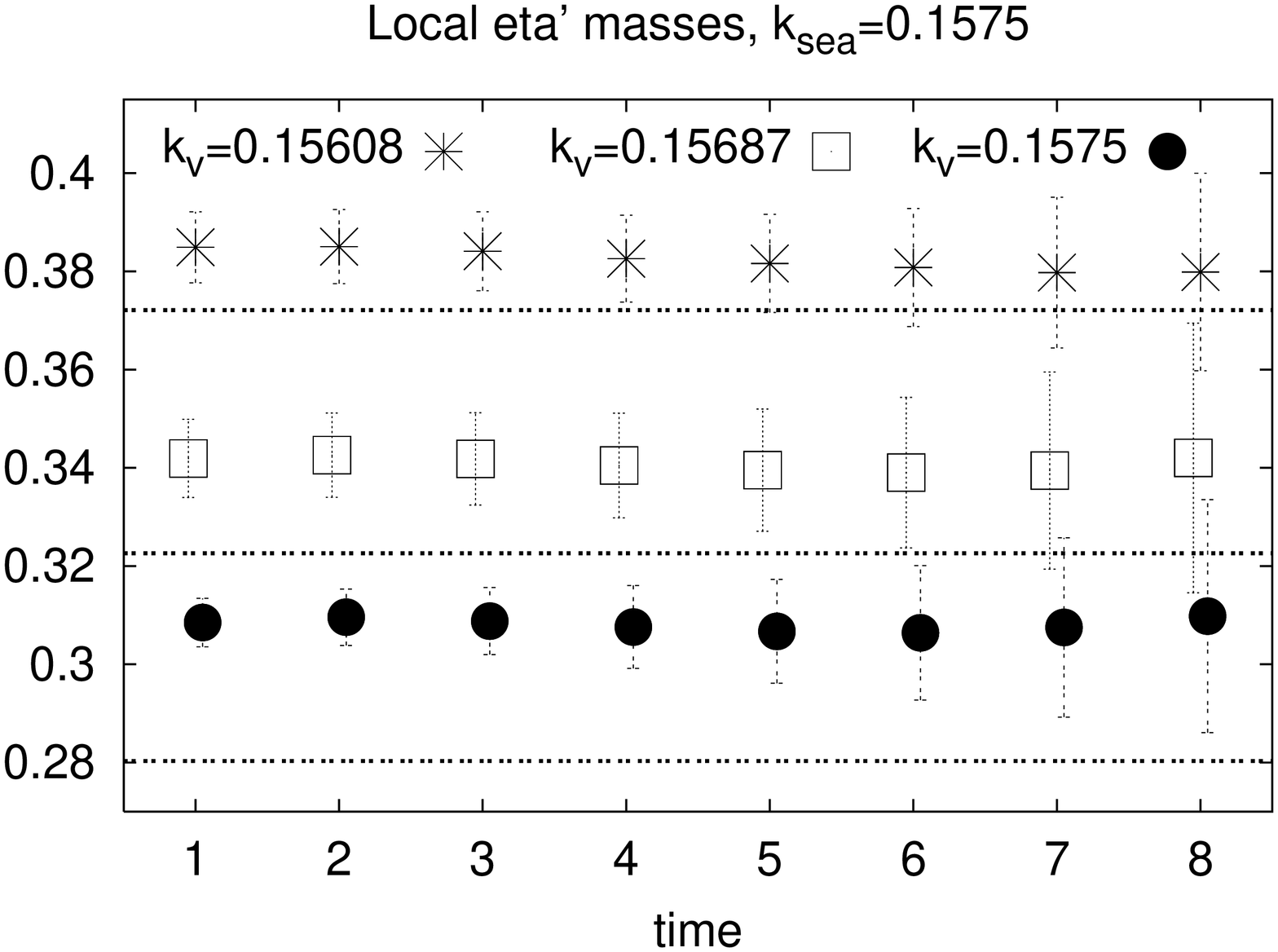}
\vskip-.7cm
\caption{
  Local $\eta'$ masses in the $u$-$d$ sector, \protect\eq{eq:udsec}, from
  PQS, for 3 values of $\kappa_{val}$.  The lines correspond to the octet
  masses.}
\label{fig:loc_mass}
\includegraphics[width=5cm,angle=270]{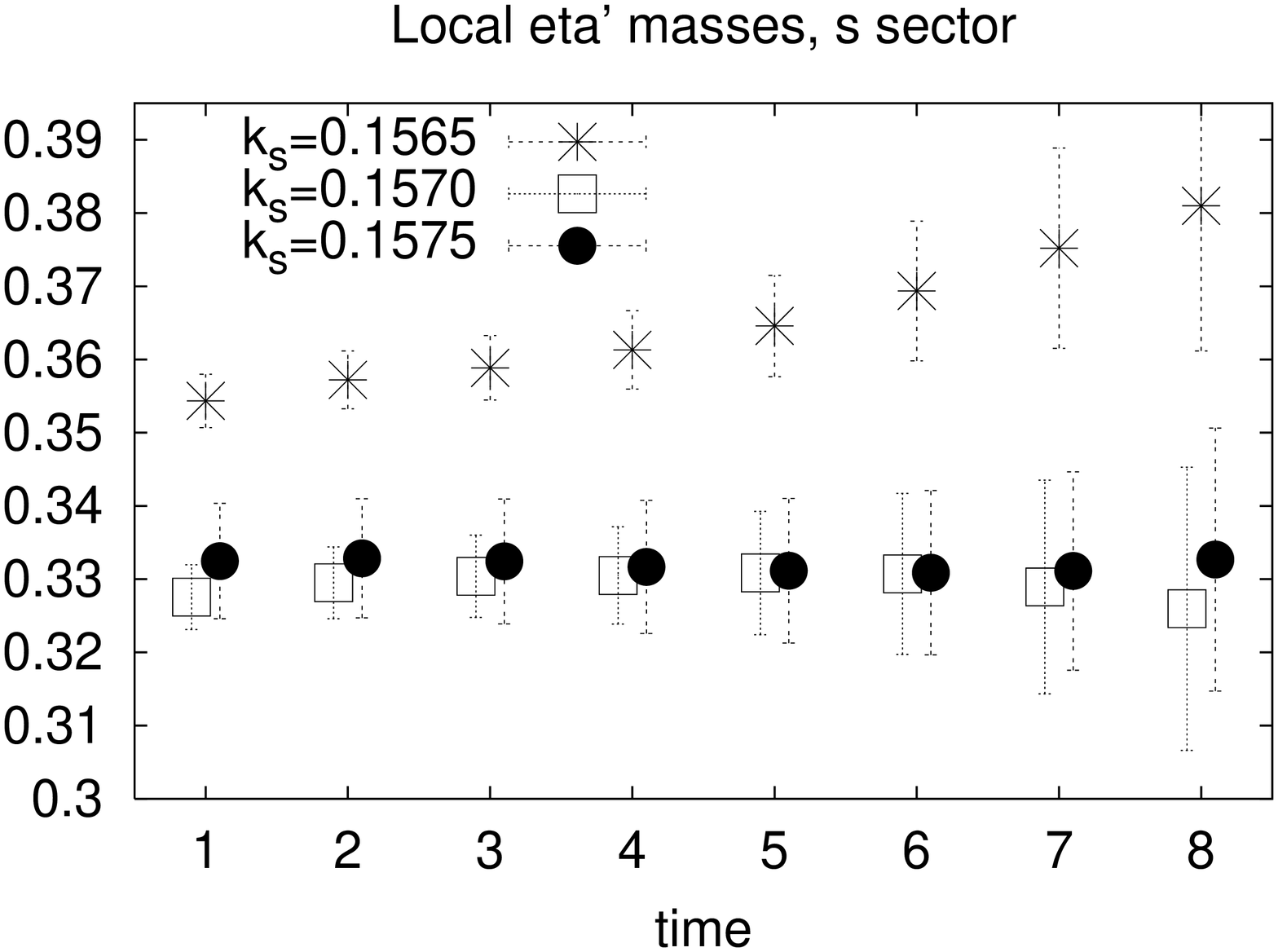}
\vskip-0.7cm
\caption{Same as \protect\fig{fig:loc_mass}
but for the  $s$ sector.}
\vskip-0.4cm
\label{fig:loc_mass_s}
\end{figure}

\section{Partially quenched $\eta - \eta'$ mass analysis}    

Given the SESAM gauge fields with just {\it two} dynamical flavors,
let us turn now to a partially quenched scenario (PQS)
\cite{Bernard:1992mk,Bernard:1994sv,Venkataraman:1997xi}: we introduce
3 quenched quarks, that will be referred to as $u$, $d$ and $s$, with
$m_u=m_d$.  For $u$ and $d$ we will choose $\kappa_{val}$,
independently from $\kappa_{sea}$. In this setting we can extrapolate
eventually to light $u$ and $d$ valence quarks at any given value of
$\kappa_{sea}$.

In PQS, the momentum space Green's functions, corresponding to the
states $ \bar{u} u, \bar{d} d$ and $ \bar{s} s$, are given by
\begin{equation}\label{eq:green}
G_{ij}=\frac{\delta_{ij}}{p^2+m_i^2}
-\frac{m_0^2}
{\left(p^2+m_i^2\right) \left(p^2+m_j^2\right) F(p^2)},
\end{equation}
with
\begin{equation}
F(p^2)= 1 +  \frac{2 m_0^2}{p^2+m_f^2} \; ,
\end{equation}
and 
\begin{eqnarray}
&2 m_0^2 =  m_{\eta'}^2- m_i^2 ,
 \;\; \eta' \equiv 
\langle  \bar{u} u + \bar{d} d \rangle \; , \label{eq:udsec}&\\
&m_0^2 =  m_{\eta'}^2- m_i^2 \; ,
 \;\; \eta' \equiv \langle \bar{s} s  \rangle \; . \label{eq:ssec}& 
\end{eqnarray}
In these formul\ae, the indices {\it i,j} refer to the valence quark
masses, while the index {\it f} points to the sea quark mass
of the QCD vacuum configurations.

For any value of time separation $t$, the mass gap, $m_0(t)$, can be
extracted by fitting the momentum zero Fourier transform of
\eq{eq:green} to the lattice data, once the local nonsinglet
pseudoscalar meson masses $m_i(t)$ have been determined from a
standard lattice analysis.  The $\eta'$ masses in the $u-d$, $s$
sectors follow subsequently from \eq{eq:udsec}, \eq{eq:ssec},
respectively.

It is nontrivial to trace the $\eta'$ signals under partial quenching,
since we see statistically well defined linear and constant terms in 
the propagators, reflecting the double pole structure. 
Their effect increase as we move $m_{val}$ away from $m_{sea}$.
To illustrate the actual $\eta'$ signal from PQS, we plotted the mass
plateaus of $\eta' \equiv \langle \bar{u} u + \bar{d} d \rangle$ for
$\kappa_{sea}=0.1575$ and three different $\kappa_{val}$ values in
\fig{fig:loc_mass}. Increasing $m_{val}$, the plateau signal remains
stable. Similarly, \fig{fig:loc_mass_s} shows the plateauing in the
$s$ sector as computed from \eq{eq:ssec}.

The quality of our plateaus gives us confidence that we will be able to
successfully complete the partially quenched calculations and move on to the
study of improved fermions.

{\bf Acknowledgments.} The computations of this study have been
carried out at NERSC, FZ-J\"ulich and on the ALiCE cluster system  at
Wuppertal University.

\vskip-1cm

\bibliographystyle{h-elsevier}
\bibliography{lit}

\end{document}